\documentclass[aps,pre,twocolumn,showpacs]{revtex4-1}
\usepackage{amsfonts}
\usepackage{bbding}
\usepackage{amsmath}
\usepackage{amsthm}
\usepackage{graphicx}
\usepackage{amssymb}
\usepackage{graphicx}
\usepackage{dcolumn}
\usepackage{bm}
\usepackage{float}
\usepackage[colorlinks,linkcolor=blue,citecolor=blue,urlcolor=blue,hyperindex,breaklinks]{hyperref}

\begin{document}
\title{Enabling the self-contained refrigerator to work beyond its limits \\by
filtering the reservoirs}
\thanks{Zi-chen He and Xin-yun Huang contributed equally to this work and
should be considered co-first authors.}
\author{Zi-chen He$^{1,2}$}
\author{Xin-yun Huang$^1$}
\author{Chang-shui Yu$^1$}
\email{ycs@dlut.edu.cn}
\affiliation{$^1$School of Physics, Dalian University of
Technology, Dalian 116024, China\\
$^2$School of Mathematical Sciences, Dalian University of
Technology, Dalian 116024, China}
\date{\today }

\begin{abstract}
 In this paper, we study the quantum self-contained refrigerator [N. Linden, S. Popescu and P. Skrzypczyk, Phys. Rev. Lett. \textbf{105}, 130401 (2010)]
in the strong internal coupling regime with engineered reservoirs. We find that if some modes of the three thermal reservoirs can be  properly filtered out, the efficiency and the working domain of the
refrigerator can be improved in contrast to the those in the weak internal coupling regime, which indicates one advantage of the strong internal coupling. In addition, we find that the background natural vacuum reservoir could cause the filtered refrigerator to stop working and the background natural thermal reservoir could greatly reduce the cooling efficiency. 
\end{abstract}

\maketitle

\section{INTRODUCTION}

Thermodynamics, as an important branch of physics, mainly deals with the properties of thermodynamic motion and the rules of the macroscopic-scale physical systems. When the physical nature is down
to the quantum level, quantum mechanics has to be taken into account in the research of the thermodynamic laws, which leads to the emergence of the quantum thermodynamics \cite{QT,modernoptics}. In the last decades, increasing interests were drawn to the 
 various quantum thermal machines \cite{gev,GAK,corre,gel,leg,prx} such as quantum engines \cite{sco,JCP,gev1,br,o3,o2,ap2,ap3,ros,science352}, quantum refrigerators \cite{geu,L3,c2,c1,ap1,howsmall} and so on.
In particular, the self-contained quantum refrigerator proposed in Ref. \cite{howsmall} reduced the complexity of a refrigerator to its extreme, and enabled the refrigerator to act as simple as a single logic gate \cite{renner}. It was even shown that the self-contained refrigerator can get as close as the Carnot efficiency \cite{PS}.
 Recently the self-contained quantum thermal machines have also
attracted wide attention in many aspects \cite{q2,q3,Bohr,sr,Br,Corre2,Mit,yu,Pat}. For example, the roles of quantum
information resources such as entanglement, quantum discord and coherence in this refrigerator have been studied, respectively, in Refs. \cite{q3,Corre2,Mit}. The thermodynamic behaviors and the
efficiency with the strong internal coupling have been addressed in Ref. 
\cite{yu}. In addition, a wide variety of other self-contained refrigerator
and autonomous thermal engine models have also been proposed \cite%
{Yixin,ven,q4,Mari,Ven,Silva,Pat}.

The self-contained refrigerator \cite{howsmall} including three
weakly-interacting qubits  only interacts with thermal reservoirs of different temperatures. No external
source of work or control such as time-dependent Hamiltonians and prescribed unitary
transformations is required.  Although the weak three-qubit interaction is one crucial dynamical mechanism to enable the refrigerator to work, the previous attempt  \cite{yu} showed that that enhancing the interaction alone between the three qubits seemed not to bring the obvious benefit to the efficiency or the working domain. Considering the wide applications of the reservoir engineering \cite{GAK,KKS,e1,e2,e3,e4,e5}, a natural question is whether the further potentiality of the self-contained refrigerator can be exploited by the properly engineered reservoirs. In addition, how does the coexistence of the engineering reservoirs and the natural background reservoirs affect the thermodynamical behaviors of the refrigerator?

\begin{figure}[tbp]
\centering
\includegraphics[width=0.8\columnwidth]{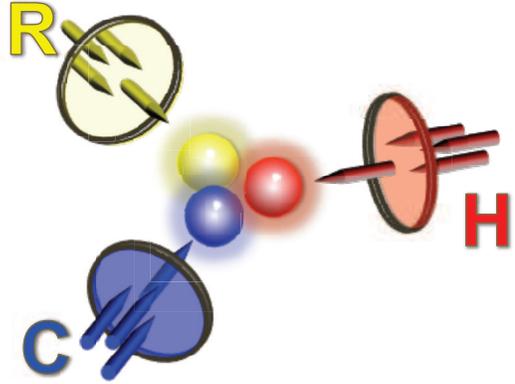}
\caption{Three coupled qubits interact separately with their reservoirs denoted by  \textit{H}, \textit{R} and \textit{C}  of different temperatures labeled by $T_H$, $T_R$ and $T_C$. Between each qubit and its reservoir is there a (maybe conceptual) filter which only allows some modes with certain frequencies in the reservoir to interact with the qubit and filers other modes out. The \textit{arrows} in the figure represent the channels by which the three qubits interact with their reservoirs. Experimentally, the filtered thermal reservoir in some cases
might be realized by a gas noise lamp and
the filter by a wave guide cutting off the certain
frequencies \cite{sco}. }
\end{figure}

In this paper, we address these questions by restudying the self-contained  refrigerator of three qubits within the strong internal coupling regime. Here we mainly take into account the engineered reservoirs with some frequency bands filtered out as well as the case with both the engineered and the natural reservoirs coexisting.  We
derive the master equation subject to the filtered reservoirs based on the
Born-Markovian approximation and study the thermal behaviors by solving the
steady-state heat currents with various filtering. It is found that in the ideal case, i.e., without the background natural reservoirs,  the refrigerator with filtering could have the higher working efficiency
 or the larger working domain than that without filtering. 
Furthermore, we show that the background natural vacuum reservoir could prohibit the heat currents to occur in the system and the background  natural thermal reservoir can greatly reduce the ability of the machine working as a refrigerator.  We would like to emphasize that the huge potential of the self-contained refrigerator is tapped by the proper application of the engineered reservoirs associated with the strong internal coupling which is the main difference from the previous jobs \cite{yu,Corre2} where no obvious advantage was shown by only enhancing the internal coupling but preserving the natural reservoirs intact. This
paper is organized as follows. In Sec. \ref{ME}, we derive the master
equation with the filtered reservoirs. In Sec. \ref{HCC}, we show the
steady-state solution of the system and analyze the thermal behaviors of the quantum
thermal machine. Finally, the conclusion is given in
Sec. \ref{CAD}.

\section{The model and the master equation}
\label{ME} 

To begin with, let's first give a detailed introduction of the
model of our refrigerator system. Our model is very similar to Refs. \cite{howsmall,PS}. As sketched in Fig. 1, this model includes
three interacting qubits, respectively,  in contact with a reservoir.  The temperature of each reservoir will be
taken to be different. We denote the temperatures of the
reservoir as $T_C$, $T_R$ and $T_H$ respectively,
which we will refer to as the ``\textit{cold}", ``\textit{room}" and ``\textit{hot}" reservoirs. Similarly, we denote the three qubits by $C$, $R$ and $H$ corresponding to the different reservoirs they are interacting with. The refrigerator means that  the heat should be extracted from the ``\textit{cold}" reservoir through qubit $C$.

The
Hamiltonian of the three free qubits reads 
\begin{equation}
H_{0}=H_{C}+H_{R}+H_{H}  \label{Equ.1}
\end{equation}%
where $H_{\alpha }=\frac{\omega _{\alpha }}{2}\sigma _{\alpha }^{z}$ with $%
\sigma _{\alpha }^{z}=\left\vert 1\right\rangle _{\alpha }\left\langle
1\right\vert -\left\vert 0\right\rangle _{\alpha }\left\langle 0\right\vert $%
, $\alpha =\{H,R,C\}$, and $\omega _{H}$, $\omega
_{C}$ and $\omega _{R}=\omega _{C}+\omega _{H}$ denoting the transition frequencies.  In addition, we set the Planck constant and the
Boltzmann's constant to be unit, i.e., $\hbar =k_{B}=1$ throughout the
paper. The interaction Hamiltonian of the three qubits is given by 
\begin{equation}
H_{HRC}=g\left( \sigma _{H}^{+}\sigma _{R}^{-}\sigma _{C}^{+}+\sigma
_{H}^{-}\sigma _{R}^{+}\sigma _{C}^{-}\right)  \label{eq1}
\end{equation}%
with $\sigma ^{+}=\left\vert 1\right\rangle \left\langle 0\right\vert $, $%
\sigma ^{-}=\left\vert 0\right\rangle \left\langle 1\right\vert $ and $g$ denoting the coupling strength.
Thus the Hamiltonian of the closed interacting system of the three qubits is 
$H_{S}=H_{0}+H_{HRC}$. This Hamiltonian $H_{S}$ can be exactly diagonalized
as 
\begin{equation}
H_{S}=\sum_{i=1}^{8}\varepsilon _{i}|\lambda _{i}\rangle \langle \lambda
_{i}|
\end{equation}%
with 
\begin{eqnarray}
\left\vert \varepsilon \right\rangle &=&\left[ \varepsilon _{1},\varepsilon
_{2},\varepsilon _{3},\varepsilon _{4},\varepsilon _{5},\varepsilon
_{6},\varepsilon _{7},\varepsilon _{8}\right] ^{T}  \notag \\
&=&\left[ \omega _{R},\omega _{H},g,-\omega _{C},\omega _{C},-g,-\omega
_{H},-\omega _{R}\right] ^{T}  \label{Equ.4}
\end{eqnarray}%
denoting the eigenvalues of $H_S$
and%
\begin{eqnarray}
&&\left\vert \lambda _{1}\right\rangle =\left\vert 111\right\rangle
;\left\vert \lambda _{2}\right\rangle =\left\vert 110\right\rangle
;\left\vert \lambda _{4}\right\rangle =\left\vert 100\right\rangle ;  \notag
\\
&&\left\vert \lambda _{5}\right\rangle =\left\vert 011\right\rangle
;\left\vert \lambda _{7}\right\rangle =\left\vert 001\right\rangle
;\left\vert \lambda _{8}\right\rangle =\left\vert 000\right\rangle ;  \notag
\\
&&\left\vert \lambda _{3}\right\rangle =\frac{\left\vert 101\right\rangle
+\left\vert 010\right\rangle }{\sqrt{2}};\left\vert \lambda
_{6}\right\rangle =\frac{\left\vert 101\right\rangle -\left\vert
010\right\rangle }{\sqrt{2}}
\end{eqnarray}%
representing the corresponding eigenvectors.
Now, let's the three qubits be, respectively, in contact with a thermal
reservoir which is described by a quantized radiation field. The
temperatures of three reservoirs are denoted by $T_{C}$, $T_{R}$ and $T_{H}$,
corresponding to the qubits \textit{C}, \textit{R} and \textit{H}. So the free Hamiltonian of the
reservoirs can be given by%
\begin{equation}
H_{\alpha }^{B}={\sum_{k}}\varpi _{k}a_{\alpha k}^{\dag
}a_{\alpha k}  \label{Equ.5}
\end{equation}%
and the Hamiltonian describing the interaction between the qubits and the thermal
reservoirs reads%
\begin{equation}
H_{SB}={\sum_{\alpha k}\phantom{}^\prime}\left( g_{\alpha k}\sigma _{\alpha
}^{-}a_{\alpha k}^{\dag }+g_{\alpha k}^{\ast }\sigma _{\alpha }^{+}a_{\alpha
k}\right) ,  \label{Equ.6}
\end{equation}%
where $g_{\alpha k}$ represents the coupling constant. 
Thus the total
Hamiltonian of the open system reads 
\begin{equation}
H=H_{S}+H_{B}+H_{SB},H_{B}=\sum\limits_{\alpha }H_{\alpha }^{B}.
\label{Equ.7}
\end{equation} 

It should be noted that the superscript prime in Eq. (\ref{Equ.6}) means that
some modes (frequency bands especially centered with some certain frequency) in the reservoirs are filtered out. Experimentally, such a filtered thermal reservoir
might be realized by a gas noise lamp and
the filter by a wave guide cutting off the certain
frequencies \cite{sco}. A similar application of the filtered reservoir can also be found in Refs. \cite{GAK}. In fact, with 
the reservoir engineering taken into account, the reservoir could be directly tailored  with the desired bath spectra by engineering. In this sense, the 'filters' in FIG. 1 are only conceptually illustrated. In addition, different physical systems that achieve the above Hamiltonian could require different reservoir engineerings \cite{KKS, e1,e2,e3,e4,e5}.

To proceed, let's derive the master equation that governs the dynamical evolution. To do so, we first expand the operators $\sigma _{\alpha }^{\pm }$ in the
picture of $H_{S}$. Thus one will obtain a series of eigen-operators $%
A_{\alpha j}\left( \omega _{\alpha j}\right) $ corresponding to $\sigma
_{\alpha }^{\pm }$ with the eigen-frequencies $\omega _{\alpha j}$ subject
to $\left[ H_{S},A\left( \omega \right) \right] =-\omega A\left( \omega
\right) $ and $A^{\dag }\left( \omega \right) =A\left( -\omega \right) $.
The concrete forms of $A_{\alpha j}\left( \omega _{\alpha j}\right) $ and
its eigen-frequencies $\omega _{\alpha j}$ are explicitly given in Appendix %
\ref{CN1}. Using these eigen-operators the Hamiltonian in the interaction
picture can be written as%
\begin{eqnarray}
H_{I}(t)&=&{\sum_{\alpha j}\phantom{}^\prime }\left[ g_{\alpha k}A_{\alpha
j}\left( \omega _{\alpha j}\right) a_{\alpha k}^{\dag }e^{-i\left( \omega
_{\alpha j}-\varpi _{\alpha k}\right) t}\right.  \notag \\
&+&\left.g_{\alpha k}^{\ast }A_{\alpha j}^{\dag }\left( \omega _{\alpha
j}\right) a_{\alpha k}e^{i\left( \omega _{\alpha j}-\varpi _{\alpha
k}\right) t}\right] .  \label{intH}
\end{eqnarray}
Following the standard derivation process and employing the Born-Markovian
approximation and the secular approximation \cite{open}, we can obtain the
master equation that governs the evolution of the density matrix $\rho $ of
the open system. The key procedure is given in Appendix \ref{CN2}.
Consequently, we arrive at 
\begin{eqnarray}
\frac{d\rho }{dt} &=&{\sum_{\alpha j}}\phantom{}^{\prime }\left[ J_{\alpha
j}^{+}\left( 2A_{\alpha j}^{\dag }\rho A_{\alpha j}-A_{\alpha j}A_{\alpha
j}^{\dag }\rho -\rho A_{\alpha j}A_{\alpha j}^{\dag }\right) \right.  \notag
\\
&+&\left. J_{\alpha j}^{-}\left( 2A_{\alpha j}\rho A_{\alpha j}^{\dag
}-A_{\alpha j}^{\dag }A_{\alpha j}\rho -\rho A_{\alpha j}^{\dag }A_{\alpha
j}\right) \right] ,  \label{master}
\end{eqnarray}%
where $J_{\alpha j}^{+}=\gamma _{\alpha j}n(\omega _{\alpha j})$ and $%
J_{\alpha j}^{-}=\gamma _{\alpha j}(1+n(\omega _{\alpha j}))$ with the mean
photon number $n\left( \omega _{\alpha j}\right) $ defined by%
\begin{equation}
n\left( \omega _{\alpha j}\right) =\frac{1}{e^{\frac{\omega _{\alpha j}}{%
T_{\alpha }}}-1},  \label{Equ.17}
\end{equation}%
and $\gamma _{\alpha }$ denoting the spontaneous emission rate (which will
be assumed to be independent of the frequency for simplicity). It is obvious
that $J_{\alpha j}^{-}(\omega _{\alpha j})=-J_{\alpha j}^{+}(-\omega
_{\alpha j})$. In addition, throughout the paper we use the subscripts to
abbreviate the corresponding frequency, e.g., $J_{\alpha j}^{\pm }=J_{\alpha
j}^{\pm }(\omega _{\alpha j})$, $n_{\alpha j}=n(\omega _{\alpha j})$ for
simplicity. The Markovian approximation requires $\gamma _{\alpha j}\ll
\left\vert \omega _{\alpha j}-\omega _{\alpha ^{\prime }j^{\prime
}}\right\vert $.

Up to now, one could have noticed that the master equation (\ref{master})
including the Hamiltonian of our system seems the same as the original
three-qubit self-contained refrigerator in the strong internal coupling
regime except the primes on the summary sign in Eq. (\ref{intH}) and Eq. (\ref%
{master}). However, we have to emphasize that the primes are just the key
points of our model. In usual, the reservoir is modeled by the infinite
harmonic oscillators with continuous frequencies from $0$ to $\infty $. Here
we use the prime to emphasize that some frequencies will be filtered out. In
this way, the quite different thermodynamic behaviors will be shown. To see the difference, let's briefly
recall one important design constraint on the refrigerator within the weak internal coupling regime (i.e., neglecting the primes and letting $g<<1$) \cite{howsmall}, that is, 
\begin{equation}
\frac{\omega _{C}}{\omega _{H}}<\frac{1-\frac{T_{R}}{T_{H}}}{\frac{T_{R}}{%
T_{C}}-1},  \label{con1}
\end{equation}%
which essentially guarantees that the heat current can flow out of the cold
reservoir and simultaneously provides an upper bound on the efficiency of
the refrigerator. As mentioned before, this constraint is also 
met in the strong internal coupling regime \cite{yu}.

Now we can turn to the case (with the primes) which we are interested in this paper. From
the eigen-operators and the eigen-frequencies given in Appendix %
\ref{CN1}, one can find that each qubit $\alpha $ interacts with
its reservoir by three channels corresponding to the frequencies $\omega
_{\alpha }$, $\omega _{\alpha }\pm g$ well separated by the strong coupling $%
g$. Suppose the frequency band centered by $\omega _{\alpha j}$ (the vicinity of $\omega _{\alpha j}$) is filtered
out. The interaction term subject to the operator $A_{\alpha j}\left( \omega
_{\alpha j}\right) $ in Eq. (\ref{intH}) can be safely neglected based on
the rotating wave approximation (one can also neglect it during the
derivation of the master equation and the result is the same). In other
words, when some frequencies $\omega _{\alpha j}$ are filtered out, we can
just let the corresponding $A_{\alpha j}\left( \omega _{\alpha j}\right) =0$
for simplicity. This means that in the ideal case the transition subject to $%
A_{\alpha j}\left( \omega _{\alpha j}\right) $ is completely idle.

\section{ Heat currents and cooling}
\label{HCC}
It has been shown that each qubit interacts with its reservoir through three
channels. Between each qubit and its reservoir, one can filter one or two channels (frequencies) out (If all the three channels are filtered out, it is equivalent to the qubit disconnected with the engineering reservoir), so there exist 6 methods to filtering the channels. Considering the three qubits involved,  there are $6^3=216$ cases corresponding to  different numbers of channels kept. If
we allow each qubit to interact with its reservoir through only one channel,
there will be 27 cases. In order to give a clear demonstration, here we mainly study the cases with only one channel left for each qubit without loss of generality. In all the cases, one can always obtain an analytic
description for the steady state and different thermodynamic behaviors will
be revealed. However, their concrete forms are very tedious. Therefore, we will study this problem in the following two cases.

\subsection{ Without the background reservoir}

As mentioned above, once some frequencies are filtered out, the
corresponding transitions will be idle, the corresponding energy levels
won't take part in the interaction, which is obviously an ideal case. In
this case, we have checked all possible filtering to keep only one channel
for each qubit. We find that 6 methods (divided into two cases)  can make the considered thermal machine achieve the function
of a refrigerator to cool the qubit \textit{C}. 

\textit{Reviving the inefficient refrigerator}.- Suppose we only keep the
channels corresponding to the frequency bands including the eigen-frequencies $\omega _{H3}=\omega _{H}+g$, $%
\omega _{R2}=\omega _{R}$, $\omega _{C1}=\omega _{C}-g$, and filter the
channels corresponding to the other eigen-frequencies out, the master equation (%
\ref{master}) becomes%
\begin{equation}
\frac{d\rho }{dt}=\sum\limits_{\alpha }\mathcal{D}_{\alpha }\left[ \rho %
\right] ,  \label{mas1}
\end{equation}%
where the dissipators $\mathcal{D}_{\alpha }\left[ \rho \right] $ are given
by
\begin{eqnarray}
D_{\alpha }\left[ \rho \right] &=&\left[ J_{\alpha j}^{+}\left( 2A_{\alpha
j}^{\dag }\rho A_{\alpha j}-A_{\alpha j}A_{\alpha j}^{\dag }\rho -\rho
A_{\alpha j}A_{\alpha j}^{\dag }\right) \right.  \label{diss1} \\
&&+\left. J_{\alpha j}^{-}\left( 2A_{\alpha j}\rho A_{\alpha j}^{\dag
}-A_{\alpha j}^{\dag }A_{\alpha j}\rho -\rho A_{\alpha j}^{\dag }A_{\alpha
j}\right) \right] _{\alpha j\in S}  \notag
\end{eqnarray}%
with $S=\left\{ H3,R2,C1\right\} $ denoting the index set.

To show the thermal behaviors of this model, we are only interested in the
steady-state case. So we have to calculate the steady state $\rho ^{S}$ of
Eq. (\ref{mas1}) by solving $\frac{d\rho ^{S}}{dt}=0$ which leads to an
eight dimensional linear equation array as%
\begin{gather}
W|\rho \rangle =0,  \label{le1} \\
\rho _{i\neq j}=0,W=\sum_{\alpha j\in S}W_{\alpha j},  \notag
\end{gather}%
where%
\begin{eqnarray*}
W_{H3} &=&\mathbb{J}_{H3}\otimes (\mathbb{I}_{+}\otimes \mathbb{I}_{-}+%
\mathbb{I}_{+}\otimes \mathbb{I}_{-}), \\
W_{R2} &=&2(\mathbb{I}_{+}\otimes \mathbb{J}_{R2}\otimes \mathbb{I}_{+}+%
\mathbb{I}_{-}\otimes \mathbb{J}_{R2}\otimes \mathbb{I}_{-}), \\
W_{C1} &=&C_{21}(\mathbb{I}_{+}\otimes \mathbb{J}_{C1}\otimes \mathbb{I}%
)C_{21}, \\
|\rho \rangle  &=&[\rho _{11}^{S},\rho _{22}^{S},\cdots ,\rho _{88}^{S}]^{T},
\end{eqnarray*}%
with $\mathbb{I=}\left( 
\begin{array}{cc}
1 & 0 \\ 
0 & 1%
\end{array}%
\right) $, $\mathbb{I}_{\pm }=\frac{\mathbb{I\pm \sigma }_{z}}{2}$, $C_{ab}$
denoting the C-NOT gate with $a$ as the control qubit, e.g. $C_{21}=\left(
\sigma _{x}\oplus \mathbb{I}\right) \otimes \mathbb{I}$ and $\mathbb{J}%
_{\alpha j}=\left( 
\begin{array}{cc}
-J_{\alpha j}^{-} & J_{\alpha j}^{+} \\ 
J_{\alpha j}^{-} & -J_{\alpha j}^{+}%
\end{array}%
\right) $. Even though this linear equation array (\ref{le1}) seems
complicated, fortunately, it is analytically solvable. Based on some
algebra, one can find Eq. (\ref{le1}) has \textit{4} different solutions the
nonzero entries of which are given as follows 
\begin{align}
(i)\text{ }\rho _{11}^{S}& =1;\text{ \ \ }(ii)\text{ }\rho _{88}^{S}=1;
\label{condition1} \\
(iii)\text{ }\left. \rho _{2i,2i}^{S}\right\vert _{i\leq 3}& =\frac{K_{i}^{+}}{%
N_{+}};(iv)\text{ }\left. \rho _{2i+1,2i+1}^{S}\right\vert _{i\leq 3}=\frac{%
K_{i}^{-}}{N_{-}};  \label{condition2}
\end{align}%
where 
\begin{eqnarray*}
K_{1\backslash 3}^{\pm } &=&2\left( J_{H3}^{\pm }+J_{C1}^{\mp }\right)
J_{R2}^{\pm }+J_{H3}^{\pm }J_{C1}^{\pm }, \\
K_{2\backslash 2}^{\pm } &=&2\left( J_{H3}^{\pm }+J_{C1}^{\mp }\right)
J_{R2}^{\mp }+J_{H3}^{\mp }J_{C1}^{\mp }, \\
K_{3\backslash 1}^{\pm } &=&2J_{R2}^{\mp }J_{C1}^{\pm }+J_{H3}^{\mp
}J_{C1}^{\pm }+2J_{H3}^{\mp }J_{R2}^{\pm },
\end{eqnarray*}%
with $N_{\pm }=\sum_{j=1}^{3}K_{j}^{\pm }$ being the normalization constant
and the subscripts $a\backslash b$ corresponding to the label for $K_{a}^{+}$
and $K_{b}^{-}$. It is obvious that the multiple steady states are of
initial-state dependence. In other words, one can control the system to
evolve to the certain branch (steady state) by choosing the proper initial
state.

With such an explicit expression of the steady-state density matrix, one can
study all the properties of the steady state of this non-equilibrium system.
Since we aim to reveal the thermodynamic behaviors of this system as a
refrigerator, we are only interested in the steady-state heat currents. 
As we know, when a system interacts with several reservoirs with the Hamiltonian of the system denoted by $H_S$, the dissipation procedure could be described by 
 many dissipators $\mathcal{D}_\alpha$ which could be connected with the same environment with different eigenfrequencies or connected with different reservoirs.  Each $\mathcal{D}_\alpha$ can be understood as a dissipation channel. In this sense, the steady-state heat current for such an open system can be defined as follows \cite{Alicki,Bou,You}. 
\begin{equation}\dot{Q}%
_{\alpha }=Tr\{H_{S}\mathcal{D}_{\alpha }\left[ \rho ^{S}\right] \},\label{hcd}\end{equation} where $\rho^S$ denotes the steady-state solution of the dissipative dynamics. $\dot{Q}
_{\alpha }$ describes the heat currents exchanged between the system and the reservoir through the given channel $\mathcal{D}_\alpha$. $\dot{Q}%
_{\alpha }>0$  means that the heat flows from the reservoir to the system, and $\dot{Q}%
_{\alpha }<0$ means that the heat flows from the system into the reservoir. 

Based on such a definition, one
can easily find that the heat current in our current model can be given by $%
\dot{Q}_{\alpha }=$ $\left\langle \varepsilon \right\vert W_{\alpha
j}\left\vert \rho \right\rangle $. Thus it is obvious that $\dot{Q}_{\alpha
}=0$ in both cases (i) and (ii) given in Eq. (\ref{condition1}), but in the
cases (iii) and (iv) given in Eq. (\ref{condition2})
\begin{eqnarray}
\dot{Q}_{C} &=&2(\omega _{C}-g)\frac{%
J_{H3}^{+}J_{R2}^{-}J_{C1}^{+}-J_{H3}^{-}J_{R2}^{+}J_{C1}^{-}}{N_{+}}, 
\notag \\
\dot{Q}_{H} &=&\frac{\omega _{H}+g}{\omega _{C}-g}\dot{Q}_{C},\dot{Q}%
_{R}=-\left( \dot{Q}_{H}+\dot{Q}_{C}\right) ,  \label{heatt}
\end{eqnarray}%
which means that both the cases (iii) and (iv) will provide the same heat
currents, even though they corresponds to the different steady states. It is
easy to see that $\dot{Q}_{H}+\dot{Q}_{C}+\dot{Q}_{R}=0$ indicating the
conservation of energy. If we only consider the three-qubit system as a
refrigerator working in the way that the heat is extracted from the cold
reservoir \textit{C} by inputting certain heat from the hot reservoir 
\textit{H}, the efficiency will be able to be understood as how much heat is
extracted by consuming certain amount of heat. A more rigorous analysis is given in Ref. \cite{PS}. Thus the efficiency $\eta _{1}
$ is given by%
\begin{equation}
\eta _{1}=\frac{\dot{Q}_{C}}{\dot{Q}_{H}}=\frac{\omega _{C}-g}{\omega _{H}+g},\label{eff}
\end{equation}%
which is obviously far less than the normally-working efficiency $\frac{%
\omega _{C}}{\omega _{H}}$ of the refrigerator without filtering, but one
will find from below that the filtering together with the strong internal
coupling can endow this thermal machine much more power than the original
refrigerator without filtering.

Note that $\dot{Q}_{\alpha }>0$ means that the heat flows out of the
reservoir, so whether the heat can be extracted from the cold reservoir
depends on whether $\dot{Q}_{C}$ is positive. Thus one can find the cold
reservoir \textit{C} can be cooled if and only if 
\begin{equation}
J_{H3}^{+}J_{R2}^{-}J_{C1}^{+}>J_{H3}^{-}J_{R2}^{+}J_{C1}^{-}.
\label{condif}
\end{equation}%
Eq. (\ref{condif}) can be rewritten as 
\begin{equation}
\frac{\omega _{C}-g}{\omega _{H}+g}<\frac{1-\frac{T_{R}}{T_{H}}}{\frac{T_{R}%
}{T_{C}}-1}.  \label{condifp}
\end{equation}%
\begin{figure}[tbp]
\centering
\includegraphics[width=0.8\columnwidth]{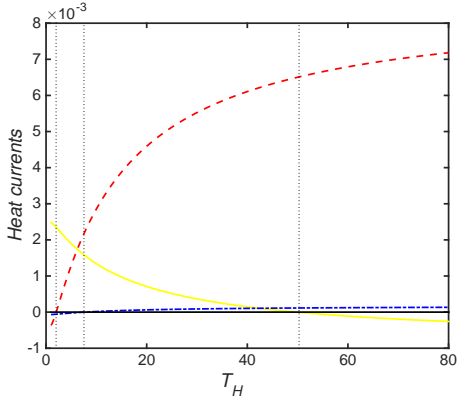}
\caption{(Color online) The heat currents $\dot{Q}_{\protect\mu }/\omega_C$ [J/s] versus $%
T_{H}$[K]. The solid yellow line, the dashed red line and the
dash-dotted blue line correspond to $\dot{Q}_{R}$, $%
\dot{Q}_{H}$, $\dot{Q}_{C}$, respectively. The horizontal solid line shows the \textit{zero} values. The vertical dotted lines are used to distinguish the four different stages defined in Eq. (\ref{stages}). With $T_H$ increasing, the four regions from left to right sequentially correspond to the four stages from Stage 1 to Stage 4.  Here $\omega _{C}=2\pi\times 210GHz$, $\omega
_{H}=3\omega_C$, $T_{C}=10K$, $T_{R}=40K$, $T_0=12K$, $\gamma =0.6\omega _{C}$.}
\end{figure} 
\begin{figure}[tbp]
\centering
\includegraphics[width=0.8\columnwidth]{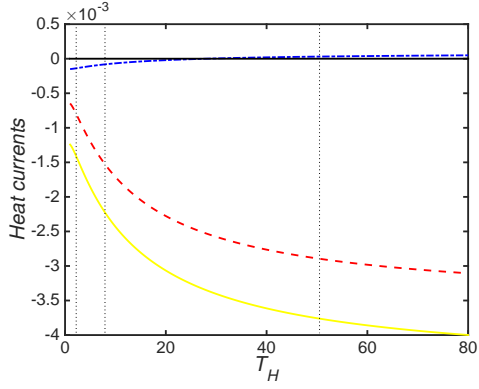}
\caption{(Color online)The heat currents $\dot{Q}^B_{\protect\mu}/\omega_C $[J/s] versus $%
T_{H}$[K]. The solid yellow line, the dashed red line and the
dash-dotted blue line correspond to $\dot{Q}^B_{R}$, $%
\dot{Q}^B_{H}$, $\dot{Q}^B_{C}$, respectively. The vertical dotted lines and all the parameters are used the same as Fig. 2.}
\end{figure}

Compared with Eq. (\ref{con1}), one can easily see that the original self-contained
refrigerator without filtering can't extract any heat from the cold
reservoir if $\frac{1-\frac{T_{R}}{T_{H}}}{\frac{T_{R}}{T_{C}}-1}<\frac{%
\omega _{C}}{\omega _{H}}$. However, if it happened that the condition in
Eq. (\ref{condif}) or Eq. (\ref{condifp}) is satisfied, we can filter some frequencies out as above
to enable the thermal machine to work normally as a refrigerator. In other
words, there exists a region $\frac{\omega _{C}-g}{\omega _{H}+g}<\frac{1-%
\frac{T_{R}}{T_{H}}}{\frac{T_{R}}{T_{C}}-1}<\frac{\omega _{C}}{\omega _{H}}$
such that the original refrigerator fails to cool the cold reservoir but our
current thermal machine can work well. Of course, one can enhance the
coupling $g$ to enlarge the working region. This is also a demonstration of
the advantage of the strong internal coupling.
We would like to emphasize that the above breakthrough in our model doesn't violate the second law of the thermodynamics. To see this, let's consider the entropy production rate in the open system. It is shown in Ref. \cite{open} that the non-equilibrium thermodynamics obeys a balance equation as \begin{equation}\sigma=\frac{dS}{dt}+J,\end{equation} where $S$ is the von Neumann entropy of the system and $J$ denotes the entropy flux due to the changes of the systematic internal energy resulting from the dissipation $\mathcal{D}(\cdot)$. $J$ is defined as \begin{equation}J=-\frac{1}{T}\left.\frac{d}{dt}\right\vert_{diss}\mathrm{Tr}\left\{H_S\rho\right\}=-\frac{1}{T}\mathrm{Tr}\left\{H_S\mathcal{D}(\rho)\right\}\end{equation}with $H_S$ and $\rho$ are the Hamiltonian and the state of the system, respectively, and the subscript means the entropy change due to the dissipative effects. The second law is guaranteed if $\sigma>0$. Considering the open system in the steady state, the density matrix of the system is independent of time, so $\frac{dS}{dt}=0$ and $\mathrm{Tr}\left\{H_S\mathcal{D}(\rho)\right\}=\sum_\alpha\frac{\dot{Q}_\alpha}{T_\alpha}$ with the heat current given in Eq. (\ref{hcd}) taken into account. Thus the second law essentially corresponds to
\begin{equation}
\sigma=-\sum_\alpha\frac{\dot{Q}_\alpha}{T_\alpha}>0.\label{secondl}
\end{equation}
Now let's turn to our current model.
From Eq. (\ref{condifp}) and Eq. (\ref{heatt}), one can easily obtain that $%
\sigma =-\frac{\dot{Q}_{H}}{T_{H}}-\frac{\dot{Q}_{C}}{T_{C}}-\frac{\dot{Q}%
_{R}}{T_{R}}>0$, which coincides with the second law of thermodynamics. 

In addition, one can easily check that the similar conclusion can also be
obtained if we only keep the bands including the eigen-frequencies $\{\omega _{H1}=\omega _{H},\omega
_{R1}=\omega _{R}-g,\omega _{C1}=\omega _{C}-g\}$ or only keep the bands including $\{\omega
_{H3}=\omega _{H}+g,\omega _{R3}=\omega _{R}+g,\omega _{C3}=\omega _{C}\}$
and filter the bands corresponding to the other eigen-frequencies out. All the details are completely
parallel, so they are omitted here.

\textit{High working efficiency}.- The above filtering method shows that our
thermal machine can work as a refrigerator in some scenario when the
original self-contained refrigerator without filtering doesn't work, but
this is not the whole story. Let's consider another filtering method, i.e.,
only keeping the channels including the eigen-frequencies $\omega _{H3}=\omega _{H}-g$, $%
\omega _{R2}=\omega _{R}$, $\omega _{C1}=\omega _{C}+g $. Repeating all the
above procedure, one can also calculate the heat currents. However, we are
interested in the efficiency which is given by 
\begin{equation}
\eta _{2}=\frac{\dot{Q}_{C}}{\dot{Q}_{H}}=\frac{\omega _{C}+g}{\omega _{H}-g}%
,  \label{efifi}
\end{equation}%
and the cooling condition which is given by 
\begin{equation}
\frac{\omega _{C}+g}{\omega _{H}-g}<\frac{1-\frac{T_{R}}{T_{H}}}{\frac{T_{R}%
}{T_{C}}-1}.  \label{region2}
\end{equation}%
It is apparent that the efficiency $\eta _{2}$ is much larger than $\frac{%
\omega _{C}}{\omega _{H}}$, but the working region given by Eq. (\ref%
{region2}) is far less than that in Eq. (\ref{con1}). This means that even though the
work region is shrunk compared with the original self-contained refrigerator
without filtering, the working efficiency can be increased.

Similarly, the large efficiency can also be achieved by only keeping the
channels corresponding to $\{\omega _{H1}=\omega _{H},\omega _{R3}=\omega _{R}+g,\omega
_{C2}=\omega _{C}+g\}$ or only keeping the channels with $\{\omega _{H2}=\omega _{H}-g,\omega
_{R2}=\omega _{R}-g,\omega _{C2}=\omega _{C}\}$.

\subsection{ With the background thermal reservoirs}

The above case is ideal, i.e., if the energy levels don't take part in the
interaction, they will be completely idle. However, in the practical
scenario, the considered three qubits \textit{H}, \textit{R} and \textit{C} cannot be isolated from their
surroundings, even though we don't impose the three reservoirs \textit{H}, \textit{R} and \textit{C} of interest. In other
words, there should exist the background thermal reservoirs which could have
the common/different temperatures closely related to the three qubits \textit{H}, \textit{R} and \textit{C}. In this sense, the
three qubits interact with their related thermal reservoirs of our interest,
at the same time, they will interact with their background thermal
reservoirs. Therefore, the master equation  (\ref{master}) should be 
\begin{equation}
\frac{d\rho }{dt}=\sum\limits_{\alpha }\phantom{}^{\prime }\mathcal{D}%
_{\alpha }\left[ \rho \right] +\sum\limits_{\beta }\mathcal{D}_{\beta }^{B}%
\left[ \rho \right]   \label{master2}
\end{equation}%
where $\sum\limits_{\alpha }\phantom{}^{\prime }\mathcal{D}_{\alpha }\left[
\rho \right] $ describes the interaction with the three filtered thermal
reservoirs of interest and $\sum\limits_{\beta }\mathcal{D}_{\beta }^{B}%
\left[ \rho \right] $ corresponds to the interaction with the background
thermal reservoirs. In order to show the effects of the background
reservoirs, we would like to study this filtered thermal machine in the
following two cases.

\textit{With the vacuum thermal reservoirs}.-In this case, we first suppose
the background reservoirs are of zero temperature. It could be a  trivial case in the practical
scenario, because the qubit \textit{C} can be directly contacted with the vacuum reservoir and the heat will spontaneously flow from the cold reservoir to the vacuum reservoir, since it can be placed in a vacuum background reservoir. After all, here we focus on the effects of the three-qubit interaction in this paper, so we requires that the three-qubit interaction isn't allowed to abandon. 

One could still intuitively
think that all the cold, the room and the hot reservoirs to be the "hot
thermal sources" and the vacuum reservoirs to be the "cold thermal
sources". The net result is that the heat would spontaneously flow from the hot thermal
sources (the three reservoirs of our interest) to the cold thermal
source (vacuum). Even if the cold qubit \textit{C} was cooled, this is
attributed to the trivial heat transport from the hot to the cold thermal
sources. Such an intuitive
understanding can be demonstrated by only keeping the channels with the eigen-frequencies $\left\{
\omega _{H2}\right. ,\omega _{R1},\left. \omega _{C3}\right\} $. The
detailed results on the heat currents are analytically given in the Appendix \ref{HC}. However, we would like to emphasize that this is not always the case. The
steady-state heat currents could be forbidden  through the system despite of the three-qubit interaction. 

To show the details, let's consider an example with only the frequencies in
the set $S=\left\{ H3,R2,C1\right\} $ kept like that in the above section.
Considering the  vacuum background reservoir, the master equation (\ref%
{master2}) can be concretely given with the specified  dissipators $\mathcal{%
D}_{\alpha }\left[ \rho \right] $ with $\alpha j\in S$ the same as Eq. (\ref%
{diss1}) and 
\begin{equation}
\mathcal{D}_{\beta }^{B}\left[ \rho \right] =\gamma _{\beta j}\left(
2A_{\beta j}\rho A_{\beta j}^{\dag }-A_{\alpha j}^{\dag }A_{\beta j}\rho
-\rho A_{\beta j}^{\dag }A_{\beta j}\right) 
\end{equation}%
with $\beta j$ covering all potential energy levels. 
\begin{figure}[tbp]
\centering
\includegraphics[width=0.8\columnwidth]{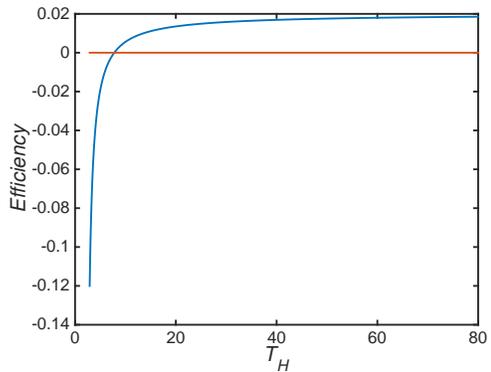}
\caption{(Color online) The efficiency versus $%
T_{H}$[K]. The negative efficiency means the different directions of heat currents $\dot{Q}_C$ and $\dot{Q}_H$. The horizontal solid line shows the value \textit{zero}.}
\end{figure}

Similarly, we need to solve the steady-state solution of Eq. (\ref{master2})
by solving $\frac{d\rho ^{S}}{dt}=0$. The corresponding linear equation
array can be given by%
\begin{equation}
\left( W-W_{\gamma }\right) |\rho \rangle =0,  \label{19}
\end{equation}%
where $W_{\gamma }=\sum_{\alpha j}W_{\alpha j}^{\gamma },$ with 
\begin{eqnarray}
W_{H1}^{\gamma } &=&2\mathbf{\Gamma }_{H1}\otimes \left( \mathbb{I}%
_{+}\otimes \mathbb{I}_{+}+\mathbb{I}_{-}\otimes \mathbb{I}_{-}\right) , \\
W_{H2}^{\gamma } &=&\mathbb{I}\otimes C_{23}\left( \mathbf{\Gamma }%
_{H2}\otimes \mathbb{I}_{-}\right) C_{23}^{\dag }, \\
W_{H3}^{\gamma } &=&\mathbf{\Gamma }_{H3}\otimes \left( \mathbb{I}%
_{+}\otimes \mathbb{I}_{-}+\mathbb{I}_{-}\otimes \mathbb{I}_{+}\right) , \\
W_{R1}^{\gamma } &=&\left( \mathbb{I}_{+}\otimes \mathbf{\Gamma }%
_{R1}\otimes \mathbb{I}_{+}+\mathbb{I}_{-}\otimes \mathbf{\Gamma }%
_{R1}\otimes \mathbb{I}_{-}\right) , \\
W_{R2}^{\gamma } &=&2\left( \mathbb{I}_{+}\otimes \mathbf{\Gamma }%
_{R2}\otimes \mathbb{I}_{-}+\mathbb{I}_{-}\otimes \mathbf{\Gamma }%
_{R2}\otimes \mathbb{I}_{+}\right)  \\
W_{R3}^{\gamma } &=&C_{13}\left( \mathbf{\Gamma }_{R3}\otimes \mathbb{I}%
\otimes \mathbb{I}_{+}\right) C_{13}^{\dagger }, \\
W_{C1}^{\gamma } &=&C_{21}\left( \mathbb{I}_{-}\otimes \mathbf{\Gamma }%
_{C1}\otimes \mathbb{I}\right) C_{21}^{\dagger } \\
W_{C2}^{\gamma } &=&\left( \mathbb{I}_{+}\otimes \mathbb{I}_{-}+\mathbb{I}%
_{-}\otimes \mathbb{I}_{+}\right) \otimes \mathbf{\Gamma }_{C2}, \\
W_{C3}^{\gamma } &=&2\left( \mathbb{I}_{+}\otimes \mathbb{I}_{+}+\mathbb{I}%
_{-}\otimes \mathbb{I}_{-}\right) \otimes \mathbf{\Gamma }_{C3},
\end{eqnarray}%
and $\mathbf{\Gamma }_{\alpha j}=\left( 
\begin{array}{cc}
-\gamma _{\alpha j} & 0 \\ 
\gamma _{\alpha j} & 0%
\end{array}%
\right) $. It is surprising that Eq. (\ref{19}) has the unique solution $%
\rho ^{S}=\left\vert 000\right\rangle \left\langle 000\right\vert $
corresponding to $\rho _{88}^{S}=1$. Thus one can easily check that no
steady-state heat current occurs between the reservoirs and the system. The
reason is that the vacuum induced spontaneous emission is much stronger than
the absorption process induced by the three thermal reservoirs \textit{H}, \textit{R} and \textit{C}.
Therefore, the system is completely decohered to the ground state when it reaches the steady state and no excitation could be induced, so the three qubits seem to be isolated with each other and no heat current could occur.

\textit{With the general thermal reservoirs}.-Now let's consider that the
three qubits \textit{H}, \textit{R} and \textit{C} are immersed in their background thermal reservoirs
with the temperature $T_{0}$ (one can also consider the case with different
temperatures, but the principle is the same). In this case, the master equation is also given by Eq. (\ref{master2}) and the dissipators $\mathcal{D}_{\alpha }\left[ \rho \right] $ are the same as Eq. (\ref{diss1})  with $\alpha j\in S=\left\{ H3,R2,C1\right\} $ and $\mathcal{D}%
_{\beta }^{B}\left[ \rho \right] $ should take the same form  as Eq. (\ref%
{master}) with $\beta j$ covering \textit{all potential energy
levels}, i.e., $\beta j=Hj,Rj,Cj, j=1,2,3$. Similarly, we need to solve steady state of the
master equation which is formally given by \begin{equation}\sum\limits_{\alpha }\phantom{}%
^{\prime }\mathcal{D}_{\alpha }\left[ \rho \right] +\sum\limits_{\beta }%
\mathcal{D}_{\beta }^{B}\left[ \rho \right] =0,\label{fis}\end{equation} 
However, the analytic
solution of Eq. (\ref{fis}) is too tedious, so we have to study the thermal behaviors via the
numerical procedure. To be clear, we use $\dot{Q}_{\alpha }=\mathrm{Tr}\{H_S\mathcal{D}_\alpha\left(\rho^S\right)\}$, as previously,
to denote the heat currents between the qubit $\alpha $ and the thermal
reservoir $\alpha $, and use $\dot{Q}_{\alpha }^{B}=\mathrm{Tr}\{H_S\mathcal{D}^B_\alpha\left(\rho^S\right)\}$ to represent the heat
current between the qubit $\alpha $ and its background thermal reservoir.
\begin{figure}[tbp]
\centering
\includegraphics[width=0.8\columnwidth]{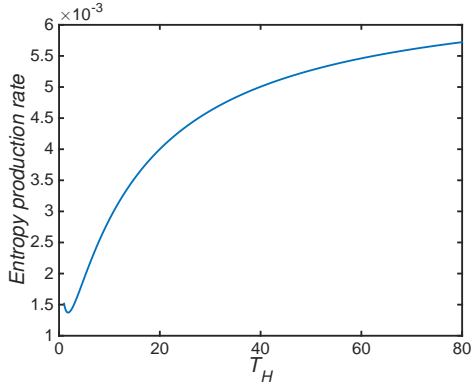}
\caption{(Color online) The entropy production rate $\sigma/\omega_C$ [J/Ks] versus $T_H$[K], where $\sigma$ is defined by $\sigma=-\sum \frac{\dot{Q}_\alpha}{T_\alpha}-\sum \frac{\dot{Q}^B_\alpha}{T_0}$. All the parameters are the same as Fig. 1.}
\end{figure}

We have plotted the heat currents $\dot{Q}_{\alpha }$ in FIG. 2  and  $\dot{Q}_{\alpha
}^{B}$  in FIG. 3 versus the temperature $T_{H}$ of the hot reservoir, where $\omega _{C}=2\pi\times 210GHz$, $\omega
_{H}=3\omega _{C},T_{C}=10K,T_{R}=40K,\gamma =0.6\omega _{C}$. We set $T_0=12K$ which is a reasonable value  compared with $T_C$ and $T_R$. It is easy to check that the condition 
given in Eq. (1) cannot be satisfied no matter what $T_H$ is. This means that
the original self-contained refrigerator without filtering cannot extract any heat from 
the cold reservoir \textit{C}. However, from Fig. 2, we can easily see that there is the positive 
heat current flowing from the reservoir \textit{C} to the qubit C which means the heat is successfully extracted. 

In order to indicates the different roles of both the engineered reservoirs and the background reservoirs, we have to analyze the behaviors of the heat currents by comparing FIG. 2 with FIG. 3. At first, one can note that in FIG. 2, the behaviors of the heat currents with the temperature $T_H$ increasing can be divided into 4 stages as follows.
\begin{equation}
\left.
\begin{array}{ll}
\mathrm{Stage\ 1}:& \{\dot{Q}_C<0,\dot{Q}_H<0,\dot{Q}_R>0\}\\  
 \mathrm{Stage\ 2}:& \{\dot{Q}_C<0,\dot{Q}_H>0,\dot{Q}_R>0\}\\ 
 \mathrm{Stage\ 3}: &\{\dot{Q}_C>0,\dot{Q}_H>0,\dot{Q}_R>0\}\\
 \mathrm{Stage\ 4}: &\{\dot{Q}_C>0,\dot{Q}_H>0,\dot{Q}_R<0\}
\end{array}  
\right.\label{stages}
\end{equation}
The four stages are marked by the vertical dotted lines both in FIG. 2 and FIG. 3.
 At stage 1, $T_H$ is the smallest, and $T_R$ is the largest. The reservoir \textit{R} serves as a hot thermal source and the others including the background reservoirs serve as the cold thermal sources. So the heat flows from the reservoir \textit{R} to the other reservoirs. Namely, from FIG. 2 one can see $\dot{Q}_R>0$ and $\dot{Q}_{H/C}<0$ and from FIG. 3 one can find $\dot{Q}^B_{H/R/C}<0$ with the subscripts $A/B$ denoting $A$ or $B$. At stage 2,  $T_H$ is slightly increased and compared with stage 1, one can find from Fig. 2 that only the direction of $\dot{Q}_H$ becomes opposite and all the others unchanged. This should be understood as a transitional period before qubit \textit{C} is cooled. At stage 3, from FIG. 2, one can see that $\dot{Q}_C>0$ which means that the heat has been extracted from the reservoir \textit{C} and the system begins to work as a refrigerator. But at this stage, the heat current  $\dot{Q}_R>0$ which is a different behavior compared with the original refrigerator with the weak internal coupling. In the regime of the weak internal coupling  \cite{howsmall,yu}, the heat is extracted from the reservoir \textit{C} and emitted into the reservoir \textit{R}. But in the current model with the background reservoir, it can be found in FIG. 3 that, $\dot{Q}^B_{H/R}<0$ which means that the heat is emitted into the background reservoirs. In particular, the background reservoir of qubit $C$ has also been cooled when $T_H$ becomes larger, which is shown by $\dot{Q}^B_C>0$ for the large $T_H$ (cf. FIG. 3). At stage 4, it is shown in FIG. 2 that $\{\dot{Q}_C>0,\dot{Q}_H>0,\dot{Q}_R<0\}$ which is very similar to the refrigerator with the weak internal coupling \cite{howsmall,yu}, and shown in FIG. 3 that $\{\dot{Q}^B_C>0,\dot{Q}^B_H<0,\dot{Q}^B_R<0\}$, which shows that the heat is extracted from both the reservoir $C$ and its relevant background reservoir \textit{C}, but the heat is emitted into the reservoir \textit{R}, the background reservoirs \textit{H} and \textit{R}.  
  Fig. 4 shows that the efficiency $\eta=\frac{\dot{Q}_C}{\dot{Q}_H}$ versus the temperature $T_H$. It is similar to the strong-internal-coupling refrigerator that the efficiency increases with the increasing $T_H$. The negative "efficiency" implies that the two heat currents $\dot{Q}_C$ and $\dot{Q}_H$ have the different directions which indicates that the qubit \textit{C} is heated. Compared with the efficiency ($\frac{\omega_C-g}{\omega_H+g}=\frac{2}{15}$) in the ideal case given in Eq. (\ref{eff}), one can find that the background reservoir greatly reduces the efficiency. As mentioned above, the validity of the second law of the thermodynamics is guaranteed by Eq. (\ref{secondl}). In the current case, one can find that the production rate of entropy is given by $\sigma=-\sum \frac{\dot{Q}_\alpha}{T_\alpha}-\sum \frac{\dot{Q}^B_\alpha}{T_0}$.  We plot $\sigma$ versus $T_H$ in FIG. 5. It is shown that $\sigma>0$, which implies the second law of the thermodynamics is satisfied.

\section{Discussion and conclusion}
\label{CAD} 
Before the end, we would like to emphasize again that  the filter could be only conceptually present. Based on the reservoir engineering, one can directly generate or simulate the reservoir with the desired spectra. In other words, the reservoir engineering per se produces the reservoir with frequency cut off instead of an ideal thermal bath \cite{GAK,e4}. So the filter has been absorbed in the engineering technique. In addition, if the filter was used, it should be heated up. This effect including its analogs generated in the reservoir engineering actually has been considered as the additional natural background reservoir.

In summary, we studied the self-contained
refrigerator composed of three qubits and the filtered thermal reservoir in the strong internal coupling regime. We have shown that the reservoir filtering can endow the self-contained refrigerator more power: The refrigerator with filtering could have the high working efficiency and the large working domain in contrast to the refrigerator without filtering. In addition, we consider the effects of the  background natural reservoirs.  The background natural vacuum reservoir in the current model could lead to that the refrigerator doesn't work (even via the heat conductivity), which could be different from what we intuitively expect. The background natural thermal reservoir can greatly reduce the cooling ability of the self-contained refrigerator in the case of the filtered reservoirs. This could shed new light on the self-contained refrigerator in the practical scenario.

\section*{ACKNOWLEDGEMENTS}
This work was supported by the National Natural Science Foundation of China,
under Grant No. 11775040 and 11375036, the Xinghai Scholar Cultivation Plan.

\bigskip \onecolumngrid
\appendix

\section{Eigen-operators of the system and Hamiltonian}

\label{CN1}

Using the eigen-vectors $\left\{ \left\vert \lambda _{i}\right\rangle
\right\} $ of the Hamiltonian $H_{S}$, the operators $\sigma _{\alpha }^{\pm
}$ can be expanded as $\sigma _{\alpha }^{-}\otimes \mathbb{I}%
=\sum\limits_{ij}\left\vert \lambda _{i}\right\rangle \left\langle \lambda
_{i}\right\vert \sigma _{\alpha }^{-}\left\vert \lambda _{j}\right\rangle
\left\langle \lambda _{j}\right\vert =\sum\limits_{\alpha j}A_{\alpha
j}\left( \omega _{\alpha j}\right) $, where $A_{\alpha j}\left( \omega
_{\alpha j}\right) $ is the eigen-operator corresponding to the
eigen-frequency $\omega _{\alpha j}$ subject to $\left[ H_{S},A\left(
\omega \right) \right] =-\omega A\left( \omega \right) $ and $A^{\dag
}\left( \omega \right) =A\left( -\omega \right) $. Concretely, $A_{\alpha
j}\left( \omega _{\alpha j}\right) $ can be explicitly given by 
\begin{eqnarray}
A_{H1} &=&\left\vert \lambda _{5}\right\rangle \left\langle \lambda
_{1}\right\vert +\left\vert \lambda _{8}\right\rangle \left\langle \lambda
_{4}\right\vert ,\omega _{H1}=\omega _{H},  \label{equ41} \\
A_{H2} &=&\frac{1}{\sqrt{2}}\left( \left\vert \lambda _{3}\right\rangle
\left\langle \lambda _{2}\right\vert +\left\vert \lambda _{7}\right\rangle
\left\langle \lambda _{6}\right\vert \right) ,\omega _{H2}=\omega _{H}-g,
\label{equ42} \\
A_{H3} &=&\frac{1}{\sqrt{2}}\left( \left\vert \lambda _{7}\right\rangle
\left\langle \lambda _{3}\right\vert -\left\vert \lambda _{6}\right\rangle
\left\langle \lambda _{2}\right\vert \right) ,\omega _{H3}=\omega _{H}+g,
\label{equ43} \\
A_{R1} &=&\frac{1}{\sqrt{2}}\left( \left\vert \lambda _{3}\right\rangle
\left\langle \lambda _{1}\right\vert -\left\vert \lambda _{8}\right\rangle
\left\langle \lambda _{6}\right\vert \right) ,\omega _{R1}=\omega _{R}-g,
\label{equ44} \\
A_{R2} &=&\left\vert \lambda _{4}\right\rangle \left\langle \lambda
_{2}\right\vert +\left\vert \lambda _{7}\right\rangle \left\langle \lambda
_{5}\right\vert ,\omega _{R2}=\omega _{R},  \label{equ46} \\
A_{R3} &=&\frac{1}{\sqrt{2}}\left( \left\vert \lambda _{8}\right\rangle
\left\langle \lambda _{3}\right\vert +\left\vert \lambda _{6}\right\rangle
\left\langle \lambda _{1}\right\vert \right) ,\omega _{R3}=\omega _{R}+g,
\label{equ47} \\
A_{C1} &=&\frac{1}{\sqrt{2}}\left( \left\vert \lambda _{3}\right\rangle
\left\langle \lambda _{5}\right\vert +\left\vert \lambda _{4}\right\rangle
\left\langle \lambda _{6}\right\vert \right) ,\omega _{C1}=\omega _{C}-g,
\label{equ48} \\
A_{C2} &=&\frac{1}{\sqrt{2}}\left( \left\vert \lambda _{4}\right\rangle
\left\langle \lambda _{3}\right\vert -\left\vert \lambda _{6}\right\rangle
\left\langle \lambda _{5}\right\vert \right) ,\omega _{C2}=\omega _{C}+g,
\label{equ49} \\
A_{C3} &=&\left\vert \lambda _{2}\right\rangle \left\langle \lambda
_{1}\right\vert +\left\vert \lambda _{8}\right\rangle \left\langle \lambda
_{7}\right\vert ,\omega _{C3}=\omega _{C}.  \label{equ50}
\end{eqnarray}%
Thus, the Hamiltonian in the interaction picture (i.e., based on the above
eigen-operators) can be given by Eq. (\ref{intH}).

\section{DERIVATION OF THE MASTER EQUATION}

\label{CN2} Following the standard procedure within the Born-Markovian
approximation \cite{open}, one can obtain the master equation as%
\begin{equation}
\frac{d\rho _{S}}{dt}=-\int\limits_{0}^{\infty
}dsTr_{B}[H_{I}(t),[H_{I}(t-s),\rho _{S}\otimes \rho _{B}]]  \label{bmm}
\end{equation}%
with $\rho _{B}=\frac{e^{-H_{H}/T_{H}}}{Z_{H}}\otimes \frac{e^{-H_{R}/T_{R}}%
}{Z_{R}}\otimes \frac{e^{-H_{C}/T_{C}}}{Z_{C}}$ with $Z_{\alpha
}=Tre^{-H_{\alpha }/T_{\alpha }}$ denoting the partition functions.
Rewriting Eq. (\ref{intH}) as 
\begin{equation}
H_{I}(t)=\sum_{\alpha j}\phantom{}^{\prime }\left[ A_{\alpha j}\left(
t\right) B_{\alpha }^{\dag }(t)+A_{\alpha j}^{\dag }\left( t\right)
B_{\alpha }(t)\right]
\end{equation}%
with $A_{\alpha j}\left( t\right) =A_{\alpha j}\left( \omega _{\alpha
j}\right) e^{-i\omega _{\alpha j}t}$ and $B_{\alpha }(t)=%
\sum_{k}\phantom{}^{\prime }g_{\alpha }a_{\alpha }e^{-i\varpi _{\alpha k}t}$
and substituting $H_{I}(t)$ given in Eq. (\ref{intH}) into Eq. (\ref{bmm}),
one can arrive at 
\begin{eqnarray}
\frac{d\rho _{S}}{dt} &=&\sum_{\alpha j}\phantom{}^{\prime }\left[
\left\langle B_{\alpha }^{\dag }(t)B_{\alpha }(t-s)\right\rangle \left(
A_{\alpha j}\rho _{S}A_{\alpha j}^{\dag }-A_{\alpha j}^{\dag }A_{\alpha
j}\rho _{S}\right) \right.  \notag \\
&&+\left. \left\langle B_{\alpha }(t)B_{\alpha }^{\dag }(t-s)\right\rangle
\left( A_{\alpha j}^{\dag }\rho _{S}A_{\alpha j}-A_{\alpha j}A_{\alpha
j}^{\dag }\rho _{S}\right) \right] +h.c.,  \label{me2}
\end{eqnarray}%
where $J_{\alpha j}^{+}=\left\langle B_{\alpha }^{\dag }(t)B_{\alpha
}(t-s)\right\rangle $ and $J_{\alpha j}^{-}=\left\langle B_{\alpha }(t)B_{\alpha }^{\dag
}(t-s)\right\rangle $ with 
\begin{equation}
\left\langle B_{\alpha }^{\dag }(t)B_{\alpha }(t-s)\right\rangle
=\int_{0}^{\infty }dse^{-i\omega _{\alpha j}s}\sum_k\phantom{}^{\prime
}\left\vert g_{\alpha k}\right\vert ^{2}Tr\left\{ \rho _{B}a_{\alpha
k}^{\dag }a_{\alpha k}\right\} .  \label{decay}
\end{equation}%
To derive Eq. (\ref{me2}), we have employed the secular approximation, i.e., 
$\left\vert \left\langle B_{\alpha }^{\dag }(t)B_{\alpha }(t-s)\right\rangle
\right\vert \ll \left\vert \omega _{\alpha }-\omega _{\alpha ^{\prime }}\
\pm g\right\vert $ with $\alpha ,\alpha ^{\prime }=R,H,C$, we can neglect
the terms with the high-frequency oscillations $e^{i\left( \omega _{\alpha
}-\omega _{\alpha ^{\prime }}\ \pm g\right) t}$. In addition, we also use
that $\left\langle B_{\alpha }^{\dag }(t)B_{\beta }(t-s)\right\rangle =0$ for $\alpha\neq\beta$.
Finally, with a reasonable spectral density $J\left( \omega \right) =%
\sum_{k}\left\vert g_{\alpha k}\right\vert ^{2}\delta \left(
\varpi _{k}-\omega \right) $, one can express $J_{\alpha j}^{\pm }=\gamma
_{\alpha j}(\pm \omega _{\alpha j})(1+n(\pm \omega _{\alpha j}))$ with an
odd function $\gamma _{\alpha j}(\pm \omega _{\alpha j})$. From $J\left(
\omega \right) $, one can easily find that $J_{\alpha j}^{\pm }$ will vanish
once $\omega $ is filtered out of the $\varpi _{k}$ due to the delta
function $\delta \left( \varpi _{k}-\omega \right) $. Consequently, there
won't be the corresponding dissipative terms to the filtered frequency in
the dissipator. This means that the transition subject to $A_{\alpha
j}\left( \omega _{\alpha j}\right) $ is completely idle. Thus one can directly let  $A_{\alpha
j}\left( \omega _{\alpha j}\right) =0$ or $\gamma _{\alpha
j}(\omega _{\alpha j})=0$ once the corresponding frequency is filtered out.

\section{Heat transport with the vacuum background reservoir}
\label{HC}
Now let's consider another case, i.e., keeping the frequencies $\left\{
\omega _{H2}\right. ,\omega _{R1},\left. \omega _{C3}\right\} $. Following
the same procedure given in the main text, one can also establish the
corresponding master equation and find the linear equation array for the
steady state. For simplicity, we let $\gamma _{\alpha j}=\gamma $
independent of the frequency. Thus the steady-state solution with the
nonzero entries can be analytically solved as 
\begin{eqnarray}
\rho _{44}^{S} &=&\frac{\rho _{66}^{S}}{2}=\frac{K}{2\left( L+N\right) +3K},
\notag \\
\rho _{77}^{S} &=&\frac{L}{K}\rho _{66}^{S},\rho _{88}^{S}=\frac{N}{K}\rho
_{66}^{S},
\end{eqnarray}%
where%
\begin{eqnarray*}
K &=&J_{H2}^{+}J_{R1}^{+}+2J_{H2}^{+}J_{C3}^{+}+2J_{R1}^{+}\tilde{J}%
_{C3}^{-}, \\
L &=&\left( J_{R1}^{+}+2J_{C3}^{+}\right) \tilde{J}_{H2}^{-}+2J_{C3}^{+}%
\left( \tilde{J}_{R1}^{-}+\gamma \right) , \\
N &=&J_{H2}^{+}\left( \tilde{J}_{R1}^{-}+\gamma \right) +2\left( \tilde{J}%
_{H2}^{-}+\tilde{J}_{R1}^{-}+\gamma \right) \tilde{J}_{C3}^{-},
\end{eqnarray*}%
with $\tilde{J}_{\alpha j}^{-}=J_{\alpha j}^{-}+\gamma $. Accordingly, the
heat currents can be given by
\begin{align}
\dot{Q}_{C}& =2\omega _{C}\gamma \frac{NJ_{C3}^{+}-LJ_{C3}^{-}}{K}\rho
_{66}^{S},  \label{hc} \\
\dot{Q}_{H}& =\left( \omega _{H}-g\right) \frac{LJ_{H2}^{+}-KJ_{H2}^{-}}{%
K}\gamma \rho _{66}^{S},  \label{hh} \\
\dot{Q}_{R}& =-\left( \omega _{R}-g\right) \frac{KJ_{R1}^{-}-NJ_{R1}^{+}%
}{K}\gamma \rho _{66}^{S},  \label{hr} \\
\dot{Q}_{H}^{B }& =-\left( 2\omega _{H}-g\right) \gamma \rho _{66}^{S},
\label{vac} \\
\dot{Q}_{C}^{B }& =-(\omega _{R}-g)\gamma \rho _{66}^{S},\dot{Q}%
_{C}^{B }=-(\frac{L+K}{K}\omega _{C}-g)\gamma \rho _{66}^{S}.
\label{vac2}
\end{align}%
Here we use $\dot{Q}_{\alpha}$ with $\alpha=\{H,R,C\}$ to denote the heat current between the system and the filtered reservoir $\alpha$ and use $\dot{Q}_{\alpha}^{B }$ to represent the heat current between the system and the background vacuum reservoir contacted with qubit $\alpha$. Thus  one can check that that $\dot{Q}_{C}+\dot{Q}_{C}^{B }+\dot{Q}%
_{H}+\dot{Q}_{H}^{B}+$ $\dot{Q}_{R}=0$ coinciding with the
conservation of energy. In particular, one can find that $\dot{Q}_{\alpha
}>0$ and $\dot{Q}_{\alpha }^{B }<0$ are always satisfied. So the
cooling effect on the qubit C is attributed to the heat transport to the
vacuum reservoirs. In addition, the heat currents $\dot{Q}_{\alpha }^{B
}$ related to the vacuum reservoirs (with zero temperature) produce infinite
positive entropy flow, which guarantees the second law of the
thermodynamics. 

\bigskip \twocolumngrid

\end{document}